# Unraveling the formation dynamics of metallic femtosecond laser induced periodic surface structures

L. Khosravi Khorashad[1], A. Reicks[1], A. Erickson[2], J. E. Shield[2], D. Alexander[1], A. Laraoui[2,3], G. Gogos[2], C. Zuhlke[1], and C. Argyropoulos[1,4,*]

[1]Department of Electrical and Computer Engineering, University of Nebraska-Lincoln, Lincoln, NE 68588, USA
[2]Department of Mechanical and Materials Engineering, University of Nebraska-Lincoln, Lincoln, NE 68588, USA
[3]Department of Physics and Astronomy and the Nebraska Center for Materials and Nanoscience, University of Nebraska-Lincoln, NE 68588, USA
[4]Department of Electrical Engineering, The Pennsylvania State University, University Park, PA 16803, USA
*cfa5361@psu.edu

## Abstract

Femtosecond laser surface processing (FLSP) is an emerging fabrication technique to efficiently control the surface morphology of many types of materials including metals. However, the theoretical understanding of the FLSP formation dynamics is not a trivial task, since it involves the interaction of various physical processes (electromagnetic, thermal, fluid dynamics) and remains relatively unexplored. In this work, we tackle this problem and present rigorous theoretical results relevant to low-fluence FLSP that accurately match the outcomes of an experimental campaign focused on the formation dynamics of laser induced periodic surface structures (LIPSS) on stainless steel. More specifically, the topography and maximum depth of LIPSS trenches are theoretically and experimentally investigated as a function of the number of laser pulses. Moreover, precise LIPSS morphology measurements are performed using atomic force microscopy (AFM). The proposed comprehensive simulation study is based on two-temperature model (TTM) non-equilibrium thermal simulations coupled with fluid dynamic computations to capture the melting metal phase occurring during FLSP. Our rigorous simulation results are found to be in excellent agreement with the AFM measurements. The presented theoretical framework to model FLSP under low-fluence femtosecond laser pulses will be beneficial to various emerging applications of LIPSS on metallic surfaces, such as cooling high-powered laser diodes and controlling the thermal emission or absorption of metals.

## Introduction

Femtosecond laser surface processing (FLSP) is an emerging method to functionalize various material surfaces at the micro and nanoscale.[1] It is a rapid fabrication technique which utilizes ultrashort laser pulses to permanently change the morphology of a material interface. The change in surface topography can occur at the nano or micron scale precision level. Owing to its high precision, fast processing, excellent repeatability, and low-cost, the FLSP method has received substantial attention regarding applications in different fields ranging from nanotechnology and medicine to engineering and science.[1–8] Due to the ultrashort femtosecond laser pulses used, FLSP leads to minimal thermal damage to the material outside the laser focal spot making it desirable property to construct different surface morphologies on temperature sensitive delicate structures.[9–11]

In previous works, during FLSP on metals, it has been demonstrated that low laser fluence illumination can lead to laser induced periodic surface structures (LIPSS).[8,12–21] However, the theoretical understanding of the LIPSS formation dynamics is not trivial, since it involves the interaction of various physical processes (electromagnetic, thermal, fluid dynamics), and remains relatively unexplored. Interestingly, many factors affect the shape formation of LIPSS, which usually depends on the laser polarization, illuminated material properties, laser fluence value, number of laser pulses, and illumination angle. Based on these factors, the surface can form different kinds of periodic nano/micron scale structures. For instance, changing the angle of laser incidence alters the periodicity of ripple formation on copper due to changes to the induced surface plasmon effect.[21] Another study demonstrated that increasing the number of pulses during the FLSP method applied to steel can transition from ripples to form deeper grooves on the surface.[12] In addition, it is possible to change the induced linear- into radial-shaped periodic ripples just by changing the linear polarization profile of the laser into a cylindrical vector beam.[22]

Hence, small changes in different factors affecting the LIPSS formation dynamics during FLSP can form different periodic structures on the material surface leading to distinct optical properties that can be of interest to various applications. Although there has been significant experimental progress in employing FLSP to create LIPSS for periodic grating-type sample fabrication, there is still a lack of comprehensive theoretical understanding on how the FLSP phenomenon leads to the LIPSS formation. The vast majority of studies in FLSP metals has been focused on predicting the period of LIPSS.[12,21,22] Very few theoretical and experimental works exist that are dedicated to the understanding of the trench depth formation dynamics that metals or dielectrics experience during LIPSS.[23,24] The scarcity in relevant studies, especially in terms of theory, is mainly because the simulation of ultrashort laser pulse interaction with bulk metallic materials is a challenging task because it involves different complex physical processes occurring at an extremely short period of time during laser illumination. Moreover, a very limited number of full-wave simulations capable of explaining all empirical observations during low-fluence FLSP exist in the literature, mainly focused on the LIPSS period prediction and not their trough depth.[12,21,22,25] Usually, the simulation method of choice highly depends on the specific goals of each research problem and the complexity of the constituent processes. Note that over the course of FLSP formation dynamics, chemical reactions, physical deformations, and mechanical distortions are present. For example, molecular dynamics can be used to investigate the material response in the extremely small molecular level and may be useful to chemical reaction simulations.[26–28] However, this modeling method cannot be applied to samples with dimensions more than several tens of nanometers due to computational constraints. On the other hand, Monte Carlo simulations can be applied to predict statistical fluctuations and stochastic propagation of light into larger sample areas but suffer from limited accuracy due to various relatively crude approximations employed.[29,30] The ultimate solution to these theoretical modeling constraints is full-wave simulations that combine multiple physics solvers, such as electromagnetic, heat, and fluid dynamics.

In this work, we develop an accurate multiphysics full-wave simulation modeling based on the finite element method to predict the temperature-dependent physical properties, heat dissipation, temperature distribution, and resulting fluid flow of the melted metallic material. We first explain the theory of heat dissipation in metals under ultrafast laser illumination by using the two-temperature model (TTM). The TTM is one of the most successful models to explain the rapid energy transfer during femtosecond laser illumination from energetic (hot) electrons to lattice vibrations by avoiding characterizing the chaotic behavior details of hot electrons at the early stage of their creation.[13,31,32] In the case of high enough laser fluence values, the metallic material experiences different structural phase transformations due to melting and re-solidification. The phase changes can include solid to liquid, liquid to vapor, and solid to vapor when the lattice temperature reaches above the melting, vaporization, and critical temperatures, respectively. We incorporate melting, vaporization (material removal), and re-solidification into our theoretical model. The computed results accurately match extensive atomic force microscopy (AFM) measurements of the depth of the resulting LIPSS trenches on stainless steel for an increased number of laser pulses. We also thoroughly discuss the important impacts to the metal surface of recoil pressure and Marangoni effect, both playing critical roles in the LIPSS formation dynamics. We envision that the presented theoretical LIPSS modeling will contribute a useful simulation tool that can be adapted to different materials (both dielectric and metal) just by changing the input material properties. It leads to an improved understanding of the LIPSS formation process that can be used to create electromagnetic absorptive (black) metallic surfaces, broadband or narrowband thermal emitters, and radiative cooling devices.[6]

## Theory and Simulations

When ultrashort laser pulses are incident on metals, their free electron gas rapidly oscillates. More specifically, the electrons are excited from their ground energy state forming non-thermal energetic electrons, also known as hot electrons.[33–43] As the time progresses, the electrons energy is relaxed via electron-electron and, subsequently, electron-phonon scattering procedures. Over the ultrafast duration of femtosecond laser pulse illumination, the temperatures of electrons and lattice are in non-equilibrium states. To accurately determine the induced heat distribution which results in deformation of the metallic surface, we need to attain a precise understanding of the electrons and lattice temperatures.

The most precise method to account for the calculation of electron and lattice temperatures under femtosecond laser illumination is the so-called two-temperature model (TTM) which consists of two coupled partial differential equations.[13,31,32] Interestingly, at the early stages of the laser-metal interaction, the electron temperature may reach up to tens of thousands of degrees. This accumulation of energy is later transferred to the lattice giving rise to the lattice temperature that causes the metal to heat up. The TTM is composed of the following Eqs. (1) and (2):

$$C_e \frac{\partial T_e}{\partial t} = \nabla \cdot (k_e \nabla T_e) - g_{e-L}(T_e - T_L) + S(\boldsymbol{r}, t), \qquad (1)$$

$$C_L \frac{\partial T_L}{\partial t} = \nabla \cdot (k_L \nabla T_L) + g_{e-L}(T_e - T_L). \tag{2}$$

Considering the index $i \equiv e$ or $L$, i.e., electron ($e$) or lattice ($L$), respectively, and $T_i$, $C_i$, and $k_i$ are the temperature, heat capacity, and thermal conductivity of electrons and lattice, respectively, where $C_L = C_{p_L} \cdot \rho_L$ with $C_{p_L}$ being the specific heat capacity of the lattice and $\rho_L$ is the lattice material density. The parameter $g_{e-L}$ in Eqs. (1) and (2) represents the electron-lattice coupling factor. The parameters of heat capacity, thermal conductivity, and coupling factor for electrons are highly temperature dependent both in terms of electron and lattice temperatures. The electron thermal conductivity can be expressed as:[31]

$$k_e = k_L \frac{BT_e}{AT_e^2 + BT_L}, \tag{3}$$

Where $A$ and $B$ are experimentally measured constants given in Table 1 that are specific to the material under study.[31] Stainless steel's main ingredient is iron (65 – 75% by weight for stainless steel 304). Hence, throughout the current work we use the physical and thermal properties of iron[12] with parameters presented in Table 1. Note that subscripts $s$ and $l$ replace subscript $L$ in the solidus and liquidus states of the lattice, respectively. Figures 1(a-c) depict the plots of electron thermal conductivity, electron heat capacity, and electron-lattice coupling factor for this material, where these parameters were obtained by using density functional theory calculations.[44]

**Table 1.** Relevant parameters of iron used in our simulations.[12,31,45]

| Symbol | Parameter | Value | Unit |
|---|---|---|---|
| $\rho_s$ | Solidus lattice density | 7.874 | g/cm$^3$ |
| $k_s$ | Solidus lattice thermal conductivity | 79.5 | W/m/K |
| $C_{p_s}$ | Solidus lattice Specific heat capacity | 475 | J/kg/K |
| $\rho_l$ | Liquidus lattice density | 6.90 | g/cm$^3$ |
| $k_l$ | Liquidus lattice thermal conductivity | 38.6 | W/m/K |
| $C_{p_l}$ | Liquidus lattice Specific heat capacity | 748 | J/kg/K |
| $A$ | Electron thermal conductivity constant | $0.98 \times 10^7$ | 1/s/K$^2$ |
| $B$ | Electron thermal conductivity constant | $2.8 \times 10^{11}$ | 1/s/K |
| $T_v$ | Vaporization temperature | 3100 | K |
| $T_m$ | Melting temperature | 1811 | K |
| $m_{iron}$ | Atomic mass | 55.845 | u |
| $L_v$ | Latent heat of vaporization | $6.088 \times 10^6$ | J/kg |
| $L_m$ | Latent heat of melting | $2.76 \times 10^5$ | J/kg |
| $\mu_l$ | Dynamic viscosity of liquidous lattice | 0.016 | $Pa \cdot s$ |
| $A_{mushy}$ | Mushy zone constant | $10^7$ | 1 |
| $T_s$ | Solidus temperature in mixed solid-liquid phase | 1786 | K |
| $T_l$ | Liquidus temperature in mixed solid-liquid phase | 1836 | K |

As a result of the laser interaction with steel, heat is deposited within a thin layer along the surface of this metallic material. The important parameter $S(\vec{r}, t)$ used as input in Eq. (1) represents the spatiotemporal laser heat source obtained in Cartesian coordinates ($r \equiv x, y, z$) and expressed as:

$$S(x, y, z, t) = C \cdot S_{temporal} \cdot S_{decay} \cdot S_{Gaussian} \cdot S_{SPP}, \quad (4)$$

where $C$ is a normalization coefficient factor defined as:

$$C = \frac{\alpha(1-R)\sqrt{4ln(2)}F}{\sqrt{\pi}\tau_p}, \quad (5)$$

with $\alpha$, $R$, and $\tau_p$, being the material (iron) absorption coefficient, reflection coefficient, and laser pulse duration, respectively, with values given in Table 2. The laser fluence value in this formula is denoted as $F$. As can be seen by Eq. (4), the heat source is defined as the product of four terms. The temporal term, $S_{temporal}$, is assumed to be Gaussian with its maximum value at $3\tau_p$ that is equal to:

$$S_{temporal} = exp\left(-4ln(2)\left(\frac{t-3\tau_p}{\tau_p}\right)^2\right). \quad (6)$$

The laser energy decays along a small depth into the metallic material (iron) and its eventual total absorption is characterized by the decay term, $S_{decay}$, in Eq. (4). If the laser-induced electromagnetic wave is incident along the negative z-axis (as schematically depicted in Fig. 3), then the decay term, $S_{decay}$, can be simply expressed as:

$$S_{decay} = exp(-\alpha|z|). \quad (7)$$

It is important to note that during the complicated process of laser interaction with the material, the topography of the surface changes during each laser pulse, mainly due to melting. As a result, the laser source interacts with a new updated surface with deeper z-axis trenches at each time step. Hence, the variable $z$ needs to be properly adjusted along the updated material interface that is always different from the initial planar interface geometry. Therefore, the depth variable $z$ needs to be dynamically modified at each simulation time step such that the correct form of the source decay term ($S_{decay}$) is properly applied at each time instance. The spatial part of the laser source, $S_{Gaussian}$ in Eq. (4), is denoted as a Gaussian function in the x-y plane with beam waist, $w$, given by the following formula:

$$S_{Gaussian} = exp\left(-\frac{x^2+y^2}{w^2}\right). \quad (8)$$

In principle, it is impossible to form surface plasmon polaritons (SPPs) on a flat metallic surface because the dispersion curves of plasmon polariton and the laser do not overlap.[46,47] In general, surfaces of polished metals always have defects and/or surface roughness which can initiate SPP creation eventually leading to ripple formation on the metal surface.[12,22] However, even if the surfaces were ideally perfectly smooth, the first pulse of laser incidence would roughen the surface. In our model, we enter the SPPs formation by using the source term, $S_{SPP}$, in Eq. (4) as:[32]

$$S_{SPP} = \left(1 + \beta \cdot \cos\left(\frac{2\pi x}{L_p}\right)\right), \tag{9}$$

where $L_p$ is the SPP spatial period with the value measured by our experiments (see Table 2), and $\beta$ is the SPP period amplitude in the range $0 \leq \beta \leq 1$. Here, we implicitly assume that the LIPSS period is approximately the same as SPP. In our calculations, we set $\beta = 0.5$, which determines the average non-uniformity of the initial energy deposition through the periodic amplitude of the induced LIPSS.[32] This results in a three-fold increase in energy deposition at the LIPSS troughs compared to their ridges, as has been explained in previous relevant works.[32]

The parameters of the laser pulse used in our simulation model are summarized in Table 2. We demonstrate the different characteristics of the laser pulse in Figure 2. Figure 2(a) shows the temporal part of the source term defined by Eq. (6) with femtosecond pulse duration $\tau_p = 35\ fs$. Figure 2(b) depicts the laser energy decay into the metal in the case of steel, and Figure 2(c) is the illustration of the spatial 2D Gaussian term along with the SPP source term with beam waist, $w = 1.05\ mm$. In this plot, we used a smaller value for the beam waist (compared to Table 2) just for illustration purposes. As it can be seen, the spatial Gaussian term acts as the envelope function for the SPP formation.

**Table 2.** Laser pulse parameters used in the simulations and experiments.

| Symbol | Parameter | Value | Unit |
|---|---|---|---|
| $\alpha$ | Absorption coefficient for iron[12] | $7.105 \times 10^5$ | 1/cm |
| $R$ | Reflection coefficient for iron[12] | 0.5 | – |
| $\tau_p$ | Laser pulse duration | 35 | fs |
| $w$ | Laser beam waist | 1.05 | $mm$ |
| $L_p$ | SPP spatial period from our experiments | 614.5 | nm |
| $F$ | Laser fluence | 0.23 | J/cm$^2$ |

Owing to the high lattice temperature induced mainly along the metal surface by the incident laser fluence, the material undergoes different phase changes leading to melting and material removal. Understanding the morphology and pattern formation dynamics resulting from phase changes within the metal and along its surface requires modeling of material flow and heat transport. The response of the material under these conditions can be conveniently described with the Navier-Stokes equations as:

$$\rho_L \frac{\partial \boldsymbol{u}}{\partial t} + \rho_L(\boldsymbol{u} \cdot \boldsymbol{\nabla})\boldsymbol{u} = \vec{\nabla} \cdot \left(-P\hat{\boldsymbol{I}} + \mu_L(\boldsymbol{\nabla}\boldsymbol{u}) + \mu_L(\boldsymbol{\nabla}\boldsymbol{u})^T\right) + \boldsymbol{F}_{Darcy}, \tag{10}$$

$$\boldsymbol{\nabla} \cdot \boldsymbol{u} = 0, \tag{11}$$

$$C_L \frac{\partial T_L}{\partial t} + C_L \boldsymbol{\nabla} \cdot (\boldsymbol{u} T_L) = \boldsymbol{\nabla} \cdot (k_L \boldsymbol{\nabla} T_L). \tag{12}$$

Equations (10)-(12) are directly related to conservation of momentum, mass, and energy, respectively. The flow of the molten steel is assumed to follow the incompressible Newtonian flow response.[48–50] The dynamic viscosity of the lattice is denoted by $\mu_L$, and the $(\dots)^T$ symbol is the matrix transpose operator. The vector variable $\boldsymbol{u}$ is the local velocity varying both spatially and temporally and $P$ is the total pressure specified as the summation of atmospheric pressure $P_{atm}$ and the recoil pressure $P_{recoil}$. The recoil pressure is approximately equal to $0.54 P_{sat}$ and the saturation pressure $P_{sat}$ is defined as:

$$P_{sat} = P_{atm} \cdot exp\left(\frac{L_v m_{iron}}{k_B T_m}\left(\frac{T_m}{T_v} - 1\right)\right), \tag{13}$$

where $m_{iron}$ and $L_v$ are the atomic mass and latent heat of iron evaporation, $k_B$ is the Boltzmann constant, and $T_m$ and $T_v$ are the iron's melting and vaporization temperatures, respectively. These parameters are provided in Table 1. Any external forces are included by the $\boldsymbol{F}_{Darcy}$ term in Eq. (10). In our model, $\boldsymbol{F}_{Darcy}$ is the Darcy damping force required to make the velocity to go to zero in the areas where the temperature is below the melting point. We will explain more about the Darcy force next when we discuss the thermal parameters of the lattice.

As mentioned earlier, the lattice temperature reaches high values which result in different phase formation in the metallic region such as, melting, vaporization, and re-solidification. Therefore, one needs to consider in the model the physical and thermal parameters of the lattice as continuous and well-defined temperature dependent functions. Hence, these properties can be defined via a liquid function expressed as:[51,52]

$$f_l = \begin{cases} 0 & T < T_s \\ \frac{T-T_s}{T_l-T_s} & T_s \leq T \leq T_l \\ 1 & T > T_l \end{cases}, \tag{14}$$

where $T_s$ and $T_l$ are the solidus and liquidus temperature states of iron in the solid-liquid mixed phase with values given in Table 1. Therefore, we can define the physical and thermal lattice material properties as:

$$k_L = f_l k_l + (1 - f_l) k_s, \tag{15}$$

$$\rho_L = f_l \rho_l + (1 - f_l) \rho_s, \tag{16}$$

$$C_{p_L} = f_l C_{p_l} + (1-f_l)C_{p_s} + L_m \left(\frac{df_l}{dT}\right), \tag{17}$$

$$\mu_L = \left(1 + (1-f_l)A_{mushy}\right)\mu_l, \tag{18}$$

where the subscript $s$ and $l$ refer to lattice solid and liquid states, respectively. In addition, $\mu_l$ is the dynamic viscosity of liquidous state of the lattice and $L_m$ is the latent heat of melting. The mushy zone constant is represented as $A_{mushy}$ in Eq. (18). All values of these parameters are given in Table 1. Having defined the liquid function in Eq. (14), we can express the Darcy force as:

$$\boldsymbol{F}_{Darcy} = -C\frac{(1-f_l)^2}{(f_l^3+b)}\boldsymbol{u}, \tag{19}$$

where $C$ in Eq. (19) is a big number ($\sim 10^6 \; kg/s/m^3$) to specify the mushy zone morphology and $b$ is a small number ($\sim 10^{-3}$) to avoid division by zero.[53] Note that the Darcy damping force is derived from the Kozeny-Carman equation and characterizes the mushy zone in all our simulations.[54,55]

So far, we have dealt with the governing equations of both laser/matter interactions (TTM model) and material fluid flow and heat transport (Navier-Stokes equations). Therefore, we must specify appropriate boundary conditions (BCs) for both coupled simulations. To reduce the computational burden time, we perform 2D simulations where only the x-z plane of the metal profile is considered, as it is schematically depicted in Figure 3. Hence, it is assumed that the material has infinite length along the y-axis, which is a valid approximation for the laser pulse interaction problem under study. This is because the beam size ($\sim 1.05 \; mm$ in Table 2) is much larger than the simulation space, so the fluence profile is relatively flat across the entire simulation space. Figure 3 shows a schematic illustration of the simulation domain where the laser is incident on the metallic surface from the top. The iron slab is shown in the 2D x-z plane. In the case of heat dissipation simulations, the surfaces a-b, a-d, and c-d have zero thermal flux BCs to mimic the domain termination along the left, right and bottom boundaries. The surface b-c is the most important area in our heating simulations since it is where the laser directly interacts with the metal and surrounding air interface. The heat flux BC for the b-c surface is defined as $h(T_v - T_L)$, where $h$ is the heat transfer coefficient that can reach very high values over the course of melting and vaporization.[56–60]

In the case of fluid dynamics simulations, the surfaces a-b, a-d, and c-d are treated as solid walls to terminate the domain along the left, right and bottom. The surface b-c is assumed to be a fluid surface under an external $P_{recoil}$ pressure applied to it. This surface undergoes melting phase change and, as a result, we define a temperature dependent surface tension given by:[50,61]

$$\sigma(T) = 1.93 - 3 \times 10^{-4} \cdot (T - T_m). \tag{20}$$

The gradient of the surface tension, with the unit $(N/m)$, causes the so-called Marangoni effect which can be considered as an additional BC applied to the b-c surface that is equal to:[50]

$$\hat{n} \cdot \left(-P\hat{I} + \mu_L(\vec{\nabla}\vec{u}) + \mu_L(\vec{\nabla}\vec{u})^T\right) = -P_{recoil}\hat{n} + \sigma(T)(\nabla_t \cdot \hat{n})\hat{n} - \nabla_t\sigma(T), \quad (21)$$

where $\nabla_t$ is the tangential gradient and $\hat{n}$ is the normal unit vector. To simulate the material removal in the system, we define a mass flux, $\dot{m}$, out of the system when the lattice temperature is greater than the vaporization temperature. The mass flux is defined as:

$$\dot{m} = h\frac{(T_l - T_v)}{L_v}, \quad (22)$$

where $h$ is the heat transfer coefficient. The same variable was used before to define the heat flux BC for the b-c surface shown in Figure 3. The mass flux defined by Eq. (22) produces a velocity that is applied to the various mesh nodes defined in our simulation model by the moving mesh method. Therefore, we specify the mesh velocity as:

$$\boldsymbol{u}_{mesh} \cdot \hat{\boldsymbol{n}} = \left(\boldsymbol{u} - \frac{\dot{m}}{\rho_L}\hat{\boldsymbol{n}}\right) \cdot \hat{\boldsymbol{n}} = \left(\boldsymbol{u} - \frac{h(T_l - T_v)}{\rho_L L_v}\hat{\boldsymbol{n}}\right) \cdot \hat{\boldsymbol{n}}. \quad (23)$$

All these aforementioned equations are used as manual inputs in our multiphysics full-wave simulations under the finite element method-based COMSOL Multiphysics platform[62] to accurately model the LIPSS formation dynamics with results presented in the next section.

## Discussion

**Single pulse theory**
We examine the results of the presented multiphysics simulations in two parts. This section is dedicated to single pulse simulation analysis which leads to a minimum trench depth on the surface that is very difficult to measure, due to its small size, and to compare with the theory. In the next section, we discuss the results of multiple pulse simulations that lead to deeper trenches and are more appropriate to be compared with experiments based on the actual LIPSS fabricated samples produced as a function of the laser pulse count. Figure 4(a) demonstrates the induced lattice temperature computed at the maximum value for a single laser pulse ($t \sim 3\tau_p = 105\ fs$), where the periodic heating distribution due to the SPP formation is clearly shown. Since the experimentally used laser focal spot ($w$) is very large compared to our simulation domain, it can be safely assumed that all the LIPSS corrugations created in our 2D simulation domain receive roughly the same amount of energy from the laser. Hence, the Gaussian part of the source, $S_{Gaussian}$, is uniformly distributed in our simulation domain.

According to the TTM theory, the electrons' energy is transferred to the material lattice giving rise to increased lattice temperature as time progresses. Figure 4(b) shows our calculations of the electron and lattice temperatures ($T_e$ and $T_L$) as a function of time at the center point of the geometry presented in Figure 4(a) (Point P). The electron temperature reaches tens of thousands of degrees in an ultrafast femtosecond time interval (duration of the pulse). These high-energy electrons (a.k.a., hot electrons) transfer their energy to the lattice at later times on the picosecond timescale. Eventually, $T_e$ and $T_L$ converge to the same value, as can be seen in Figure 4(b).

The remaining results presented in Figures 4(c)-4(e) are focused on the zoomed in area in the middle part of Figure 4(a). These results provide a better understanding of the underlying physics under single laser pulse illumination. More specifically, the colormap in Figure 4(c) demonstrates the lattice temperature distribution at a later stage of the time-domain simulation ($t = 1300 fs$), where the induced velocity vectors are also depicted in the same figure as white arrows. The arrows are proportional to the magnitude of the local velocity at each point. The material is removed from the system only at high lattice temperature values where increased velocity is present. The direction of the velocity vectors is a direct indication of the competition between the recoil pressure and Marangoni effect. The recoil pressure exerts force along the melted surface and pushes the material from the LIPSS trenches to the ridges. The Marangoni effect also happens at the same time because there is a surface tension gradient due to the temperature dependent nature of surface tension. The surface effects of the recoil pressure and Marangoni effect are more clearly demonstrated in Figure 4(e) where the colormap of the velocity values for the same time snap as in Figure 4(c) at $t = 1300 fs$ is presented. At the center of the LIPSS trench, the material is pushed down with higher velocity fields. Similarly, the material is expelled to the LIPSS ridges which introduces higher velocity vector fields at each side of the LIPSS trench. Finally, Figure 4(d) shows the created LIPSS at the end of the first pulse when the system reaches thermal equilibrium (or $t \to \infty$), i.e., room temperature, as it is obvious from the temperature colormap. We have defined $t \to \infty$ as the time it takes for the domain to reach room temperature ($T_0 = 300\ K$). In our model, $t \to \infty$ is taken at $100\ ns$. It is important to mention that at large pulse duration (nanosecond or millisecond) the competition between the recoil pressure and Marangoni effect has enough time to push the melted material away from the trench to the ridges during the LIPPS formation process. This results in an elevated morphology above the initial flat metallic surface even at low laser intensity. However, the presented femtosecond pulse laser excitation occurs in an extremely short time duration for the Marangoni effect and recoil pressure to take the full effect, although extremely small elevations are still obtained in the geometry but are not visible in current results.

**Multiple pulse theory**
The laser repetition rate used in our experiments is $50\ Hz$. However, it is extremely computationally intensive to consider the $\frac{1}{50} Hz = 20\ ms$ duration between each laser pulse to be able to model multiple pulses considering the fact that the pulse duration is on the order of femtoseconds. Note that the time is long enough between each pulse for the system to reach room temperature before the start of the next pulse. Therefore, in our simulations, we consider the start of consecutive laser pulses to be when the system reaches room temperature, i.e., thermal steady state ($t \to \infty$), which happens at much shorter time scales than the 20 ms spacing between pulses for the experimentally used repetition rate. This adjustment makes our multiphysics simulation possible for higher number of laser pulses while accurately corresponding to the appropriate experimental conditions. Figure 5(a) depicts 2D views of the LIPSS formation dynamics when different number of laser pulses are applied to the system but always plotted at the end of the last pulse in each view. These results are obtained at room temperature (end of pulses) when there is no longer any material removal, and the system has reached its thermal steady state with zero material velocity. The maximum indentation depth into the material after ten pulses is about $95\ nm$. The indentation depth becomes $112\ nm$, $132\ nm$, $168\ nm$, and $208\ nm$ after fifteen, twenty, thirty, and forty pulses, respectively. Figure 5(b) depicts a 3D "top" view elongated towards the y-axis of the simulated LIPSS formation at the maximum induced temperature of the 10$^{th}$ laser pulse, i.e., not at thermal steady state. Additionally, Figure 5(c) illustrates the 3D velocity distribution map at the peak temperature of the 10$^{th}$ laser pulse under the same view angle used in Figure 5(b). Again, the traces of the competition between recoil pressure at the trench center and Marangoni effect at the ridges are clearly depicted due to the higher material velocity fields in these regions.

**Experiments and theory comparison**

Next, we perform experiments with the goal of verifying the theoretical results. Eight mirror finished (average roughness of $122\ nm$) stainless steel 304 (mainly composed of iron 65 – 75% by weight) samples were illuminated by a Ti:sapphire femtosecond laser (Coherent Legend) delivering pulses with a duration of 35 fs and 800 nm central wavelength. The Gaussian beam was focused to a 1/e$^2$ diameter of 1.05 mm in all of the experiments. Single spots were illuminated with the pulse energy fixed at $1,010\ \mu J$, resulting in a peak laser fluence of 0.23 J/cm$^2$. The laser repetition rate was reduced from 1 kHz to 50 Hz using a phase-locked chopping wheel. The total number of pulses was controlled using a fast mechanical shutter with an open and close time faster than the 20 ms spacing between pulses. Spots were created in pulse count increments of 10 ranging from 10 to 60. Scanning electron microscopy (SEM) images taken at the center of the resulting craters on the sample surface are shown in Figure 6. In particular, Figure 6(a) demonstrates that after 10 pulses ripples start to faintly appear over the entire metallic surface. Between 20 and 30 pulses, the ripples start growing deeper and taking on the initial LIPSS shape. For pulse counts greater than 30 (Figure 6(d - f)), the ripples appear fully formed and only continue to increase in depth for additional number of pulses. The LIPSS period is measured from the SEM images to be equal to 614.5 ± 39.0 nm, which coincides with the excited SPP period along this metallic surface as was shown in various previous works.[12,22,32]

The small size of the LIPSS corrugations makes them difficult to measure using most surface roughness analysis techniques. Therefore, an atomic force microscope (AFM, Bruker Innova) was used to accurately measure the depth of the ripple trenches in order to compare the experimental results with the simulations. The AFM scans were taken for an area of 10 μm × 10 μm. An example of an AFM scan is included in Figure 6(i) for the case of 20 laser pulses. The goal of the currently presented work is to provide an accurate theoretical model to explain the experimental LIPSS results. Towards this goal, we compute the depth of the resulted LIPSS trenches by using our simulation approach and compare the results with the experimentally derived values taken from AFM measurements. Figure 7 is the resulting plot of the LIPSS trench maximum depth computed by theory (red dots) and AFM experiments (black dots) as a function of number of pulses in increments of ten pulses when the laser fluence value is always fixed to $0.23\ J/cm^2$. The simulation results are in excellent agreement with AFM measurements. It should be noted that the stainless-steel alloy 304 sample used in the experiments has an average roughness of $122\ nm$ before being laser processed. Hence, we always subtract this unavoidable roughness from the AFM measurements of the fabricated LIPSS trench average maximum depth to achieve a fair comparison with theory. The results presented in Figure 7 provide direct proof that the developed full-wave multiphysics theoretical model is accurate and can precisely predict the experimentally obtained results. The modulation depth increases an average of 30.8 nm for every 10 pulses, or 3.08 nm per pulse, which also agrees to our simulation results depicted in Figure 5(a). These type of AFM measurements offer valuable insights into the dynamics of FLSP formation in metals, especially when combined with the currently presented multiphysics theoretical modeling.

Finally, it is worth mentioning that the presented theoretical model has been realized by using frequency domain electromagnetic wave simulations and the time dependent laser source is introduced only in the thermal analysis. A more rigorous model will incorporate fully time dependent electromagnetic wave computations efficiently coupled to the already time dependent heat dissipation calculations. Furthermore, the current model is restricted to pulse counts less than one hundred owing to the high computational cost and increased calculation time required to perform the presented modeling. Further work to realize efficient simulations with pulse counts greater than one hundred will require different computational modelling approaches than the current one, such as the level-set method[63] where there is no moving mesh incorporation and, as a result, the simulations become much faster. Moreover, in this work, we did not study the roughness and other defects on the initial flat metallic surface. The study of roughness is of essential importance to explain the initial SPP formation dynamics and will be the subject of our future work.

## Conclusions

We have fabricated various LIPSS samples on stainless steel 304 with different laser pulse counts and a fixed laser fluence. For each sample, the average value of the resulting maximum LIPSS trench depth has been measured by using the accurate AFM method applied to metallic LIPSS for the first time in the literature to our knowledge. The experimental results are verified by a new multiphysics simulation-based modeling approach. In our theoretical framework, we incorporate the two-temperature model for the simulation of the complicated non-equilibrium heat dynamics induced by the ultrashort femtosecond laser illumination of metals. Computational fluid dynamics modeling is also incorporated in our theoretical approach to accurately simulate the ultrashort laser pulse interaction with metal for high number of pulses resulting to deformations in the metallic surface and LIPSS creation. The results of the presented full-wave multiphysics simulations are in excellent agreement with the AFM-measured LIPSS morphology data. More specifically, the simulations can accurately predict the maximum LIPSS trench depth for increasing number of incident laser pulses. The currently presented theoretical method is general and can be applied to different materials (metals, semiconductors, dielectrics) just by changing the parameters introduced in the model. It can be a very useful theoretical tool to predict the LIPSS formation dynamics that can be used in various applications, such as electromagnetic absorbers and radiative cooling thermal control devices.[64,65]

## Acknowledgement

This material is based upon research supported in part by the U.S. Office of Naval Research under award numbers N00014-19-1-2384 and N00014-20-1-2025. The research was performed in part in the Nebraska Nanoscale Facility: National Nanotechnology Coordinated Infrastructure supported by the National Science Foundation under award no. ECCS-1542182, and with support from the Nebraska Research Initiative through the Nebraska Center for Materials and Nanoscience and the Nanoengineering Research Core Facility at the University of Nebraska-Lincoln. Additional partial support by the NSF/EPSCoR RII Track-1: Emergent Quantum Materials and Technologies (EQUATE), Award OIA-2044049 and NSF DMR-2224456 is also acknowledged.

# References

(1) Vorobyev, A. Y.; Guo, C. Direct Femtosecond Laser Surface Nano/Microstructuring and Its Applications. *Laser & Photonics Reviews* **2013**, *7* (3), 385–407. https://doi.org/10.1002/lpor.201200017.
(2) Papadopoulou, E. L.; Samara, A.; Barberoglou, M.; Manousaki, A.; Pagakis, S. N.; Anastasiadou, E.; Fotakis, C.; Stratakis, E. Silicon Scaffolds Promoting Three-Dimensional Neuronal Web of Cytoplasmic Processes. *Tissue Engineering Part C: Methods* **2010**, *16* (3), 497–502. https://doi.org/10.1089/ten.tec.2009.0216.
(3) Zuhlke, C. A.; Anderson, T. P.; Li, P.; Lucis, M. J.; Roth, N.; Shield, J. E.; Terry, B.; Alexander, D. R. Superhydrophobic Metallic Surfaces Functionalized via Femtosecond Laser Surface Processing for Long Term Air Film Retention When Submerged in Liquid. In *Laser-based Micro- and Nanoprocessing IX*; SPIE, 2015; Vol. 9351, pp 102–111. https://doi.org/10.1117/12.2079164.
(4) Peng, E.; Roth, A.; Zuhlke, C. A.; Azadehranjbar, S.; Alexander, D. R.; Gogos, G.; Shield, J. E. 3D Electron Microscopy Characterization of Ag Mound-like Surface Structures Made by Femtosecond Laser Surface Processing. *Applied Surface Science* **2019**, *480*, 1047–1053. https://doi.org/10.1016/j.apsusc.2019.02.197.
(5) Kruse, C.; Tsubaki, A.; Zuhlke, C.; Alexander, D.; Anderson, M.; Peng, E.; Shield, J.; Ndao, S.; Gogos, G. Influence of Copper Oxide on Femtosecond Laser Surface Processed Copper Pool Boiling Heat Transfer Surfaces. *Journal of Heat Transfer* **2019**, *141* (5). https://doi.org/10.1115/1.4043129.
(6) Reicks, A.; Tsubaki, A.; Anderson, M.; Wieseler, J.; Khorashad, L. K.; Shield, J. E.; Gogos, G.; Alexander, D.; Argyropoulos, C.; Zuhlke, C. Near-Unity Broadband Omnidirectional Emissivity via Femtosecond Laser Surface Processing. *Commun Mater* **2021**, *2* (1), 1–11. https://doi.org/10.1038/s43246-021-00139-w.
(7) Tsubaki, A.; Anderson, M.; Reicks, A.; Shield, J. E.; Alexander, D. R.; Zuhlke, C. A. Multi-Material, Multi-Layer Femtosecond Laser Surface Processing. In *Laser-based Micro- and Nanoprocessing XV*; SPIE, 2021; Vol. 11674, pp 17–30. https://doi.org/10.1117/12.2582756.
(8) *Optically Induced Nanostructures: Biomedical and Technical Applications*; König, K., Ostendorf, A., Eds.; De Gruyter: Berlin, 2015.
(9) Seitz, B.; Langenbucher, A.; Hofmann-Rummelt, C.; Schlötzer-Schrehardt, U.; Naumann, G. O. H. Nonmechanical Posterior Lamellar Keratoplasty Using the Femtosecond Laser (Femto-Plak) for Corneal Endothelial Decompensation. *American Journal of Ophthalmology* **2003**, *136* (4), 769–772. https://doi.org/10.1016/S0002-9394(03)00449-5.
(10) Tsibidis, G. D.; Mansour, D.; Stratakis, E. Damage Threshold Evaluation of Thin Metallic Films Exposed to Femtosecond Laser Pulses: The Role of Material Thickness. *Optics & Laser Technology* **2022**, *156*, 108484. https://doi.org/10.1016/j.optlastec.2022.108484.
(11) Lagunov, V. L.; Rybachuk, M.; Itthagarun, A.; Walsh, L. J.; George, R. Modification of Dental Enamel, Dentin by an Ultra-Fast Femtosecond Laser Irradiation: A Systematic Review. *Optics & Laser Technology* **2022**, *155*, 108439. https://doi.org/10.1016/j.optlastec.2022.108439.
(12) Tsibidis, G. D.; Mimidis, A.; Skoulas, E.; Kirner, S. V.; Krüger, J.; Bonse, J.; Stratakis, E. Modelling Periodic Structure Formation on 100Cr6 Steel after Irradiation with Femtosecond-Pulsed Laser Beams. *Appl. Phys. A* **2017**, *124* (1), 27. https://doi.org/10.1007/s00339-017-1443-y.
(13) Tsibidis, G. D.; Barberoglou, M.; Loukakos, P. A.; Stratakis, E.; Fotakis, C. Dynamics of Ripple Formation on Silicon Surfaces by Ultrashort Laser Pulses in Subablation Conditions. *Phys. Rev. B* **2012**, *86* (11), 115316. https://doi.org/10.1103/PhysRevB.86.115316.
(14) Stratakis, E.; Bonse, J.; Heitz, J.; Siegel, J.; Tsibidis, G. D.; Skoulas, E.; Papadopoulos, A.; Mimidis, A.; Joel, A.-C.; Comanns, P.; Krüger, J.; Florian, C.; Fuentes-Edfuf, Y.; Solis, J.; Baumgartner, W. Laser Engineering of Biomimetic Surfaces. *Materials Science and Engineering: R: Reports* **2020**, *141*, 100562. https://doi.org/10.1016/j.mser.2020.100562.

(15) Tsibidis, G. D.; Stratakis, E.; Loukakos, P. A.; Fotakis, C. Controlled Ultrashort-Pulse Laser-Induced Ripple Formation on Semiconductors. *Appl. Phys. A* **2014**, *114* (1), 57–68. https://doi.org/10.1007/s00339-013-8113-5.
(16) Tsibidis, G. D.; Fotakis, C.; Stratakis, E. From Ripples to Spikes: A Hydrodynamical Mechanism to Interpret Femtosecond Laser-Induced Self-Assembled Structures. *Phys. Rev. B* **2015**, *92* (4), 041405. https://doi.org/10.1103/PhysRevB.92.041405.
(17) Guosheng, Z.; Fauchet, P. M.; Siegman, A. E. Growth of Spontaneous Periodic Surface Structures on Solids during Laser Illumination. *Phys. Rev. B* **1982**, *26* (10), 5366–5381. https://doi.org/10.1103/PhysRevB.26.5366.
(18) Zuhlke, C. A.; Anderson, T. P.; Alexander, D. R. Fundamentals of Layered Nanoparticle Covered Pyramidal Structures Formed on Nickel during Femtosecond Laser Surface Interactions. *Applied Surface Science* **2013**, *283*, 648–653. https://doi.org/10.1016/j.apsusc.2013.07.002.
(19) Choi, J.; Choi, W.; Shin, Y.-G.; Han, S.; Kim, K.-S.; Cho, S.-H. Enhancement Periodic Regularity of Surface Nano Ripple Structures on Si Wafer Using a Square Shaped Flat-Top Beam Femtosecond NIR Laser. *Appl. Phys. A* **2021**, *128* (1), 46. https://doi.org/10.1007/s00339-021-05144-x.
(20) Wang, H.; Pöhl, F.; Yan, K.; Decker, P.; Gurevich, E. L.; Ostendorf, A. Effects of Femtosecond Laser Shock Peening in Distilled Water on the Surface Characterizations of NiTi Shape Memory Alloy. *Applied Surface Science* **2019**, *471*, 869–877. https://doi.org/10.1016/j.apsusc.2018.12.087.
(21) Zuhlke, C. A.; Tsibidis, G. D.; Anderson, T.; Stratakis, E.; Gogos, G.; Alexander, D. R. Investigation of Femtosecond Laser Induced Ripple Formation on Copper for Varying Incident Angle. *AIP Advances* **2018**, *8* (1), 015212. https://doi.org/10.1063/1.5020029.
(22) Tsibidis, G. D.; Skoulas, E.; Stratakis, E. Ripple Formation on Nickel Irradiated with Radially Polarized Femtosecond Beams. *Opt. Lett., OL* **2015**, *40* (22), 5172–5175. https://doi.org/10.1364/OL.40.005172.
(23) Bonse, J.; Rosenfeld, A.; Krüger, J. On the Role of Surface Plasmon Polaritons in the Formation of Laser-Induced Periodic Surface Structures upon Irradiation of Silicon by Femtosecond-Laser Pulses. *Journal of Applied Physics* **2009**, *106* (10), 104910. https://doi.org/10.1063/1.3261734.
(24) Chang, C.-L.; Cheng, C.-W.; Chen, J.-K. Femtosecond Laser-Induced Periodic Surface Structures of Copper: Experimental and Modeling Comparison. *Applied Surface Science* **2019**, *469*, 904–910. https://doi.org/10.1016/j.apsusc.2018.11.059.
(25) Fraggelakis, F.; Tsibidis, G. D.; Stratakis, E. Ultrashort Pulsed Laser Induced Complex Surface Structures Generated by Tailoring the Melt Hydrodynamics. *OEA* **2022**, *5* (3), 210052–210052. https://doi.org/10.29026/oea.2022.210052.
(26) Shugaev, M. V.; Gnilitskyi, I.; Bulgakova, N. M.; Zhigilei, L. V. Mechanism of Single-Pulse Ablative Generation of Laser-Induced Periodic Surface Structures. *Phys. Rev. B* **2017**, *96* (20), 205429. https://doi.org/10.1103/PhysRevB.96.205429.
(27) Lipp, V.; Ziaja, B. Classical Molecular Dynamics Simulations of Surface Modifications Triggered by a Femtosecond Laser Pulse. *Modelling* **2022**, *3* (3), 333–343. https://doi.org/10.3390/modelling3030021.
(28) Yamashita, Y.; Yokomine, T.; Ebara, S.; Shimizu, A. Heat Transport Analysis for Femtosecond Laser Ablation with Molecular Dynamics-Two Temperature Model Method. *Fusion Engineering and Design* **2006**, *81* (8), 1695–1700. https://doi.org/10.1016/j.fusengdes.2005.09.011.
(29) Kirner, S. V.; Wirth, T.; Sturm, H.; Krüger, J.; Bonse, J. Nanometer-Resolved Chemical Analyses of Femtosecond Laser-Induced Periodic Surface Structures on Titanium. *Journal of Applied Physics* **2017**, *122* (10), 104901. https://doi.org/10.1063/1.4993128.
(30) Florian, C.; Déziel, J.-L.; Kirner, S. V.; Siegel, J.; Bonse, J. The Role of the Laser-Induced Oxide Layer in the Formation of Laser-Induced Periodic Surface Structures. *Nanomaterials* **2020**, *10* (1), 147. https://doi.org/10.3390/nano10010147.
(31) Chen, A. M.; Xu, H. F.; Jiang, Y. F.; Sui, L. Z.; Ding, D. J.; Liu, H.; Jin, M. X. Modeling of Femtosecond Laser Damage Threshold on the Two-Layer Metal Films. *Applied Surface Science* **2010**, *257* (5), 1678–1683. https://doi.org/10.1016/j.apsusc.2010.08.122.

(32) Wang, J.; Guo, C. Numerical Study of Ultrafast Dynamics of Femtosecond Laser-Induced Periodic Surface Structure Formation on Noble Metals. *Journal of Applied Physics* **2007**, *102* (5), 053522. https://doi.org/10.1063/1.2776004.

(33) Khosravi Khorashad, L.; Argyropoulos, C. Unraveling the Temperature Dynamics and Hot Electron Generation in Tunable Gap-Plasmon Metasurface Absorbers. *Nanophotonics* **2022**, *11* (17), 4037–4052. https://doi.org/10.1515/nanoph-2022-0048.

(34) Brongersma, M. L.; Halas, N. J.; Nordlander, P. Plasmon-Induced Hot Carrier Science and Technology. *Nature Nanotechnology* **2015**, *10* (1), 25–34. https://doi.org/10.1038/nnano.2014.311.

(35) Hartland, G. V. Optical Studies of Dynamics in Noble Metal Nanostructures. *Chem. Rev.* **2011**, *111* (6), 3858–3887.

(36) Kale, M. J.; Christopher, P. Plasmons at the Interface. *Science* **2015**, *349* (6248), 587–588. https://doi.org/10.1126/science.aac8522.

(37) Wu, K.; Chen, J.; McBride, J. R.; Lian, T. Efficient Hot-Electron Transfer by a Plasmon-Induced Interfacial Charge-Transfer Transition. *Science* **2015**, *349* (6248), 632–635. https://doi.org/10.1126/science.aac5443.

(38) Furube, A.; Hashimoto, S. Insight into Plasmonic Hot-Electron Transfer and Plasmon Molecular Drive: New Dimensions in Energy Conversion and Nanofabrication. *NPG Asia Materials* **2017**, *9* (12), e454–e454. https://doi.org/10.1038/am.2017.191.

(39) Govorov, A. O.; Zhang, H.; Gun'ko, Y. K. Theory of Photoinjection of Hot Plasmonic Carriers from Metal Nanostructures into Semiconductors and Surface Molecules. *J. Phys. Chem. C* **2013**, *117* (32), 16616–16631. https://doi.org/10.1021/jp405430m.

(40) Zhang, H.; Govorov, A. O. Optical Generation of Hot Plasmonic Carriers in Metal Nanocrystals: The Effects of Shape and Field Enhancement. *J. Phys. Chem. C* **2014**, *118* (14), 7606–7614. https://doi.org/10.1021/jp500009k.

(41) Kong, X.-T.; Wang, Z.; Govorov, A. O. Plasmonic Nanostars with Hot Spots for Efficient Generation of Hot Electrons under Solar Illumination. *Advanced Optical Materials* **2017**, *5* (15).

(42) Khosravi Khorashad, L.; Besteiro, L. V.; Correa-Duarte, M. A.; Burger, S.; Wang, Z. M.; Govorov, A. O. Hot Electrons Generated in Chiral Plasmonic Nanocrystals as a Mechanism for Surface Photochemistry and Chiral Growth. *J. Am. Chem. Soc.* **2020**, *142* (9), 4193–4205. https://doi.org/10.1021/jacs.9b11124.

(43) Laraoui, A.; Halté, V.; Vomir, M.; Vénuat, J.; Albrecht, M.; Beaurepaire, E.; Bigot, J.-Y. Ultrafast Spin Dynamics of an Individual CoPt3 Ferromagnetic Dot. *Eur. Phys. J. D* **2007**, *43* (1), 251–253. https://doi.org/10.1140/epjd/e2007-00120-y.

(44) Lin, Z.; Zhigilei, L. V.; Celli, V. Electron-Phonon Coupling and Electron Heat Capacity of Metals under Conditions of Strong Electron-Phonon Nonequilibrium. *Phys. Rev. B* **2008**, *77* (7), 075133. https://doi.org/10.1103/PhysRevB.77.075133.

(45) Semak, V.; Matsunawa, A. The Role of Recoil Pressure in Energy Balance during Laser Materials Processing. *J. Phys. D: Appl. Phys.* **1997**, *30* (18), 2541. https://doi.org/10.1088/0022-3727/30/18/008.

(46) Raether, H. Surface Plasmons on Smooth Surfaces. In *Surface Plasmons on Smooth and Rough Surfaces and on Gratings*; Raether, H., Ed.; Springer Tracts in Modern Physics; Springer: Berlin, Heidelberg, 1988; pp 4–39. https://doi.org/10.1007/BFb0048319.

(47) Maier, S. A. *Plasmonics: Fundamentals and Applications*; Springer US: New York, NY, 2007. https://doi.org/10.1007/0-387-37825-1.

(48) Sharma, S.; Mandal, V.; Ramakrishna, S. A.; Ramkumar, J. Numerical Simulation of Melt Pool Oscillations and Protuberance in Pulsed Laser Micro Melting of SS304 for Surface Texturing Applications. *Journal of Manufacturing Processes* **2019**, *39*, 282–294. https://doi.org/10.1016/j.jmapro.2019.02.022.

(49) Dal, M.; Fabbro, R. [INVITED] An Overview of the State of Art in Laser Welding Simulation. *Optics & Laser Technology* **2016**, *78*, 2–14. https://doi.org/10.1016/j.optlastec.2015.09.015.

(50) Shen, H.; Pan, Y.; Zhou, J.; Yao, Z. Forming Mechanism of Bump Shape in Pulsed Laser Melting of Stainless Steel. *Journal of Heat Transfer* **2017**, *139* (6). https://doi.org/10.1115/1.4035710.

(51) Zhou, C.; Deng, H.; Chen, G.; Zhang, Y.; Wang, D.; Zhou, X. Numerical Simulation of Single-Pulse Laser Ablation for Dressing a Bronze-Bond Diamond Grinding Wheel. *Precision Engineering* **2016**, *43*, 78–85. https://doi.org/10.1016/j.precisioneng.2015.06.012.

(52) Wang, R.; Lei, Y.; Shi, Y. Numerical Simulation of Transient Temperature Field during Laser Keyhole Welding of 304 Stainless Steel Sheet. *Optics & Laser Technology* **2011**, *43* (4), 870–873. https://doi.org/10.1016/j.optlastec.2010.10.007.

(53) Courtois, M.; Carin, M.; Masson, P. L.; Gaied, S.; Balabane, M. A New Approach to Compute Multi-Reflections of Laser Beam in a Keyhole for Heat Transfer and Fluid Flow Modelling in Laser Welding. *J. Phys. D: Appl. Phys.* **2013**, *46* (50), 505305. https://doi.org/10.1088/0022-3727/46/50/505305.

(54) He, X.; Fuerschbach, P. W.; DebRoy, T. Heat Transfer and Fluid Flow during Laser Spot Welding of 304 Stainless Steel. *J. Phys. D: Appl. Phys.* **2003**, *36* (12), 1388. https://doi.org/10.1088/0022-3727/36/12/306.

(55) Ki, H.; Mazumder, J.; Mohanty, P. S. Modeling of Laser Keyhole Welding: Part I. Mathematical Modeling, Numerical Methodology, Role of Recoil Pressure, Multiple Reflections, and Free Surface Evolution. *Metall Mater Trans A* **2002**, *33* (6), 1817–1830. https://doi.org/10.1007/s11661-002-0190-6.

(56) Liu, M.; Ma, G.; Zhang, X.; Zheng, D. Numerical Simulation on the Melting Kinetics of Steel Scrap in Iron-Carbon Bath. *Case Studies in Thermal Engineering* **2022**, *34*, 101995. https://doi.org/10.1016/j.csite.2022.101995.

(57) Li, J.; Brooks, G. A.; South, C.; Provatas, N. Phase-Field Modeling of Steel Scrap Melting in a Liquid Steel Bath. **2004**.

(58) Kruskopf, A. A Model for Scrap Melting in Steel Converter. *Metall Mater Trans B* **2015**, *46* (3), 1195–1206. https://doi.org/10.1007/s11663-015-0320-3.

(59) Penz, F. M.; Tavares, R. P.; Weiss, C.; Schenk, J.; Ammer, R.; Pastucha, K.; Klösch, G. Analytical and Numerical Determination of the Heat Transfer Coefficient between Scrap and Hot Metal Based on Small-Scale Experiments. *International Journal of Heat and Mass Transfer* **2019**, *138*, 640–646. https://doi.org/10.1016/j.ijheatmasstransfer.2019.04.085.

(60) Chevallier, E. C.; Bruyère, V.; Bernard, G.; Namy, P. Femto-Second Laser Texturing Prediction Using COMSOL Multiphysics. In *COMSOL Conference*; COMSOL Technical Papers and Presentations: Boston, 2018.

(61) Sahoo, P.; Debroy, T.; McNallan, M. J. Surface Tension of Binary Metal—Surface Active Solute Systems under Conditions Relevant to Welding Metallurgy. *Metall Mater Trans B* **1988**, *19* (3), 483–491. https://doi.org/10.1007/BF02657748.

(62) COMSOL - Software for Multiphysics Simulation. www.comsol.com.

(63) Courtois, M.; Carin, M.; Le Masson, P.; Gaied, S.; Balabane, M. Complete Heat and Fluid Flow Modeling of Keyhole Formation and Collapse during Spot Laser Welding. *ICALEO* **2013**, *2013* (1), 77–84. https://doi.org/10.2351/1.5062966.

(64) Butler, A.; Schulz, J.; Argyropoulos, C. Tunable Directional Filter for Mid-Infrared Optical Transmission Switching. *Optics Express* **2022**, *30* (22), 39716–39724. https://doi.org/10.1364/OE.474728.

(65) Butler, A.; Argyropoulos, C. Mechanically Tunable Radiative Cooling for Adaptive Thermal Control. *Applied Thermal Engineering* **2022**, *211*, 118527. https://doi.org/10.1016/j.applthermaleng.2022.118527.

# Figures

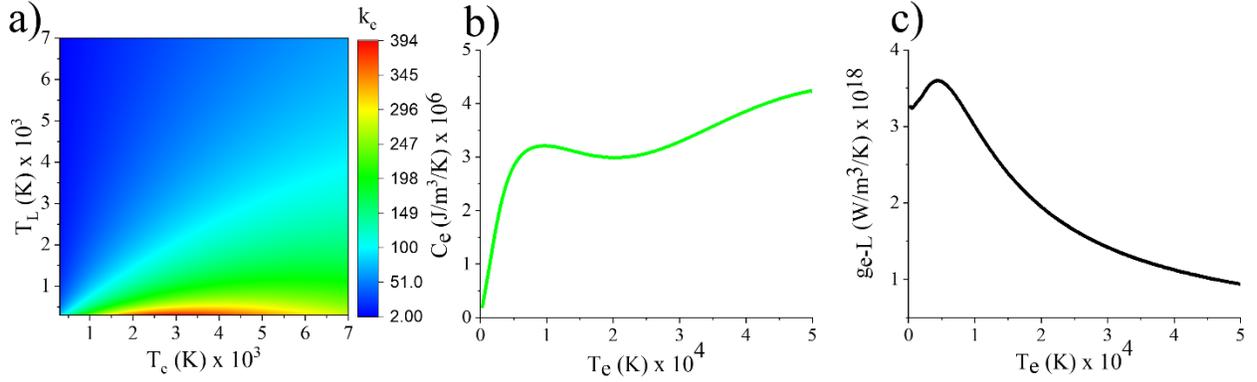

**Figure 1.** Iron parameters of (a) electron thermal conductivity as a function of electron and lattice temperatures in W/m/K, (b) electron heat capacity and (c) electron-lattice coupling factor as a function of electron temperature.

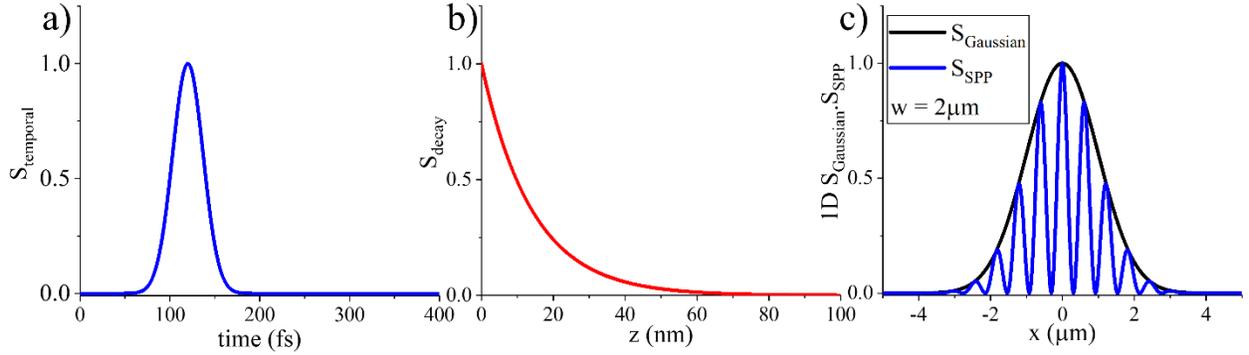

**Figure 2.** (a) Temporal part of the source. (b) Decay term of the source plotted along the z-axis (depth in metal). (c) SPP term plotted in x-direction only along with the x-direction spatial Gaussian part acting as the envelope function. We have chosen small beam waist, $w = 2\ \mu m$ for illustration purposes and all quantities are normalized. In our actual experiments, the beam waist is given in Table 2.

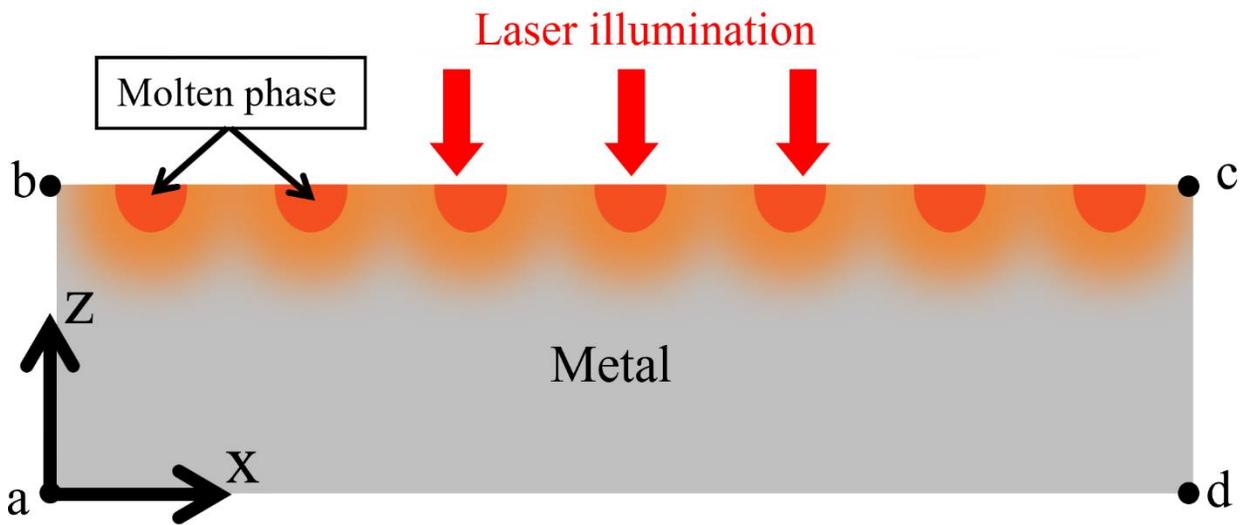

**Figure 3.** Schematic representation in the 2D x-z plane of a femtosecond laser beam incident on metal (iron) accompanied by molten phase creation. The boundaries are the following: a-b, b-c, c-d, and a-d.

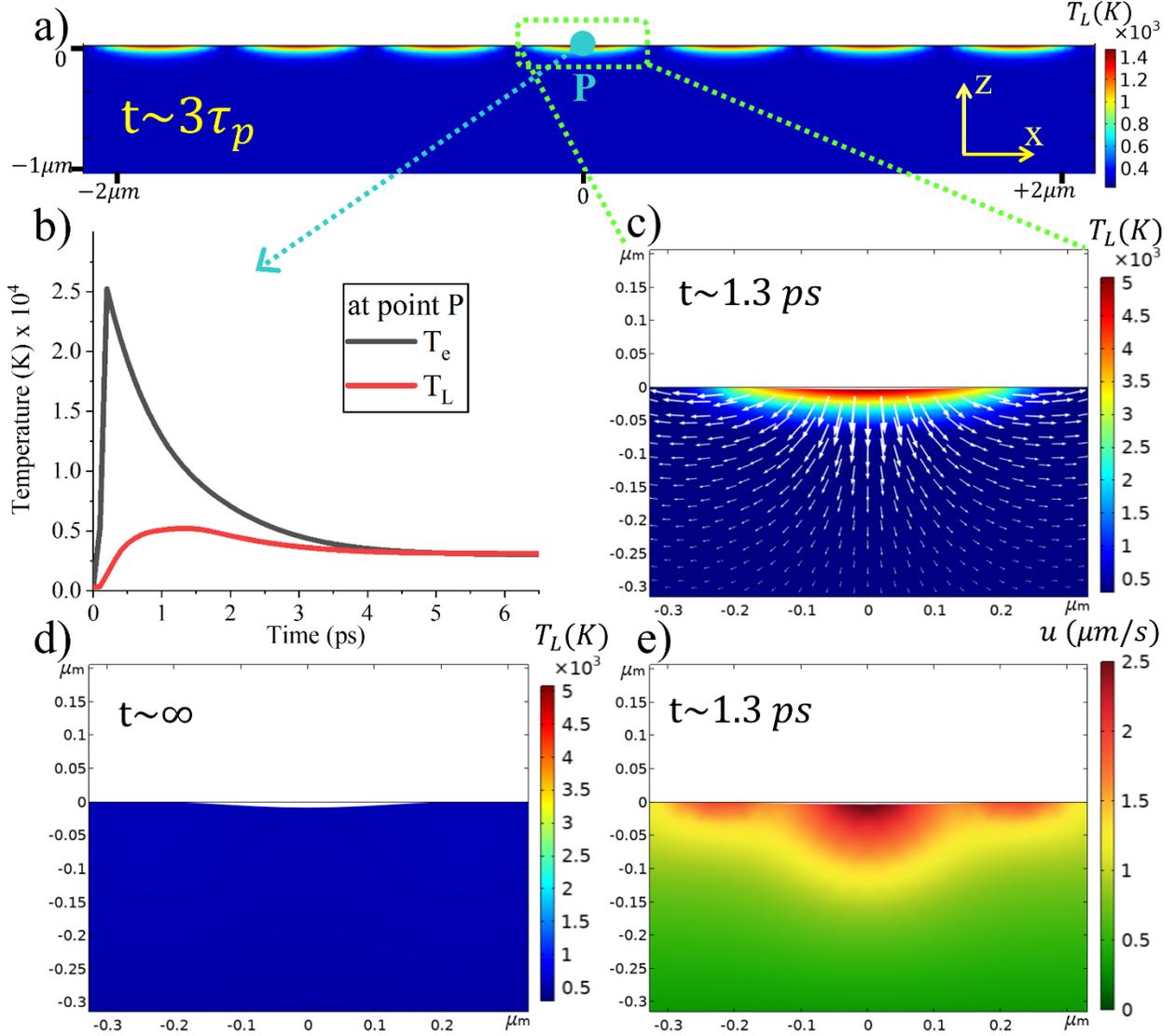

**Figure 4.** (a) Induced lattice temperature distribution at the laser pulse maximum value time $t = 3\tau_p$ observed in x-z plane. (b) Computed electron and lattice temperatures ($T_e$ and $T_L$) as a function of time monitored at Point P of Figure 4(a). (c) Induced lattice temperature distribution in the zoomed area of LIPSS at a later stage in the time-domain simulation ($t = 1300 fs$). The white arrows are the induced velocity vectors. (d) Temperature distribution and surface profile of the maximum depth trench when the sample has reached thermal steady state, i.e., room temperature. (e) Velocity magnitude distribution monitored at the same time as (c) ($t = 1300 fs$).

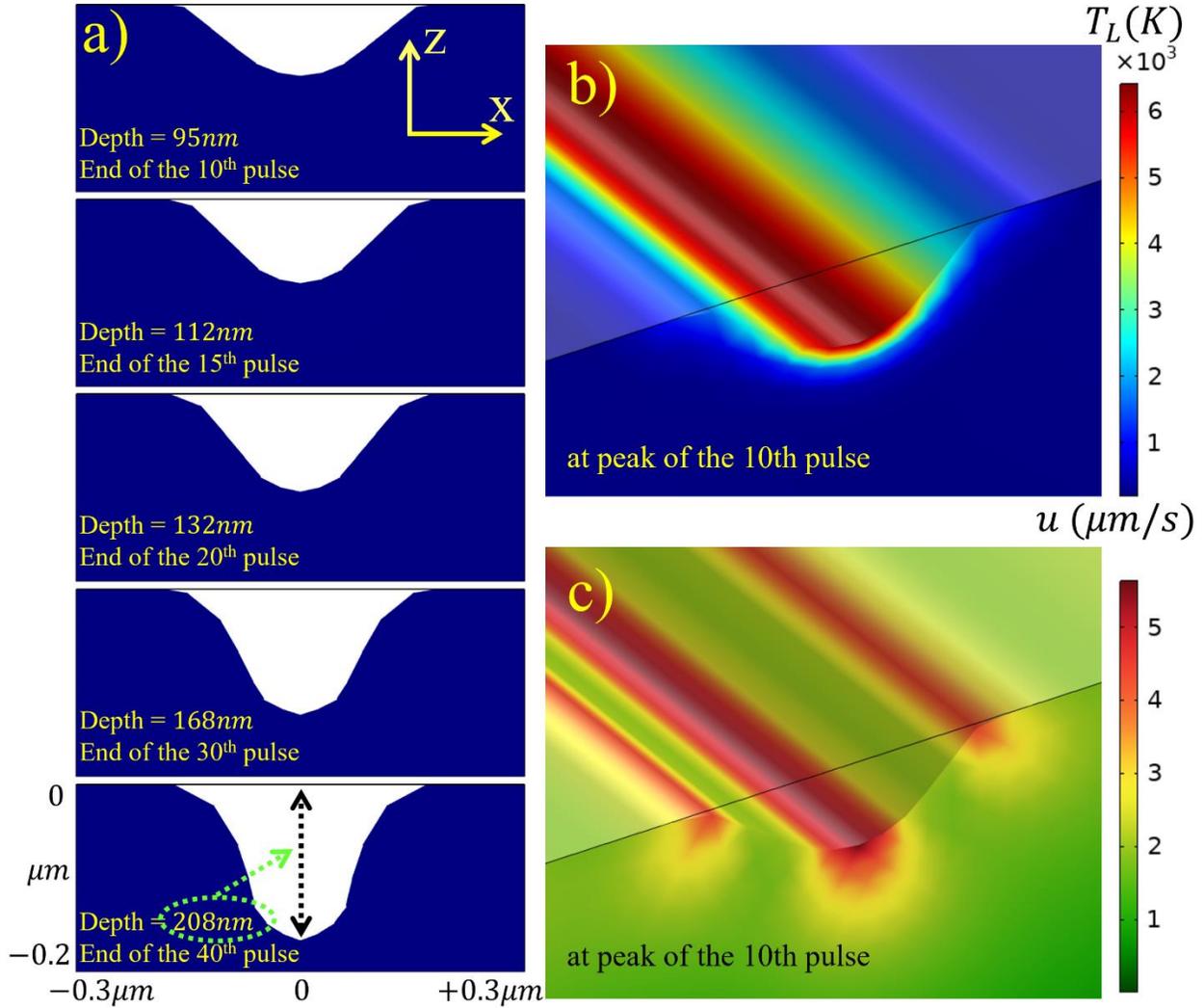

**Figure 5.** (a) Snapshots of the central LIPSS area induced by 10, 15, 20, 30, and 40 pulses plotted in the x-z plane. These results are obtained at room temperature when thermal steady state is reached. (b) Lattice temperature and (c) material velocity 3D distribution maps plotted at the peak temperature of the 10$^{th}$ pulse (not thermal steady state).

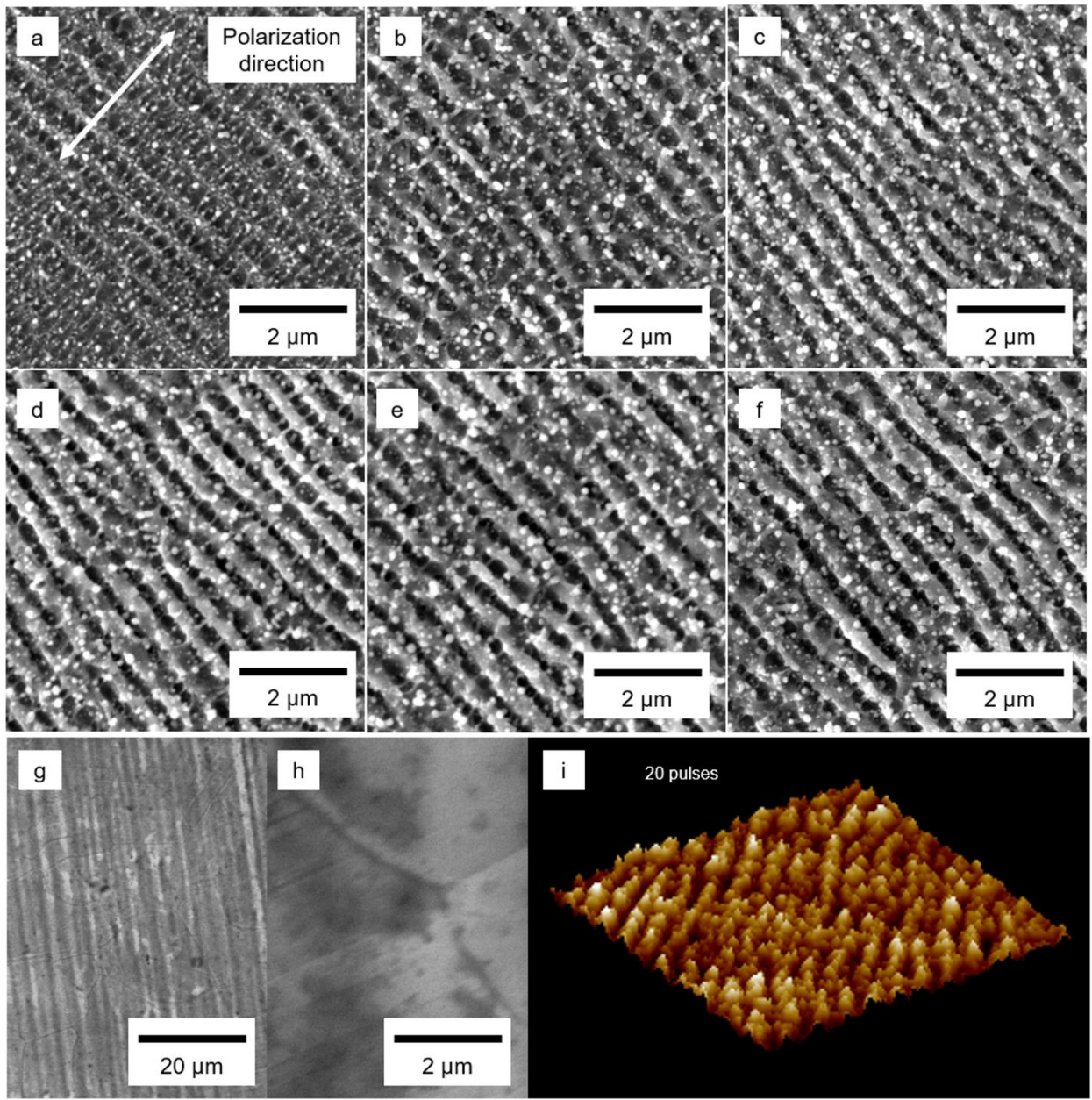

**Figure 6.** (a)-(f) SEM images of LIPSS produced with varying number of laser pulses: (a) 10, (b) 20, (c) 30, (d) 40, (e) 50, and (f) 60. The polarization of the incident laser pulse for (a) – (f) is indicated by the double arrow in (a). (g)-(h) SEM images of the unprocessed material. (i) AFM scan of LIPSS formation after 20 laser pulses. The scan size in (i) is 10 μm × 10 μm.

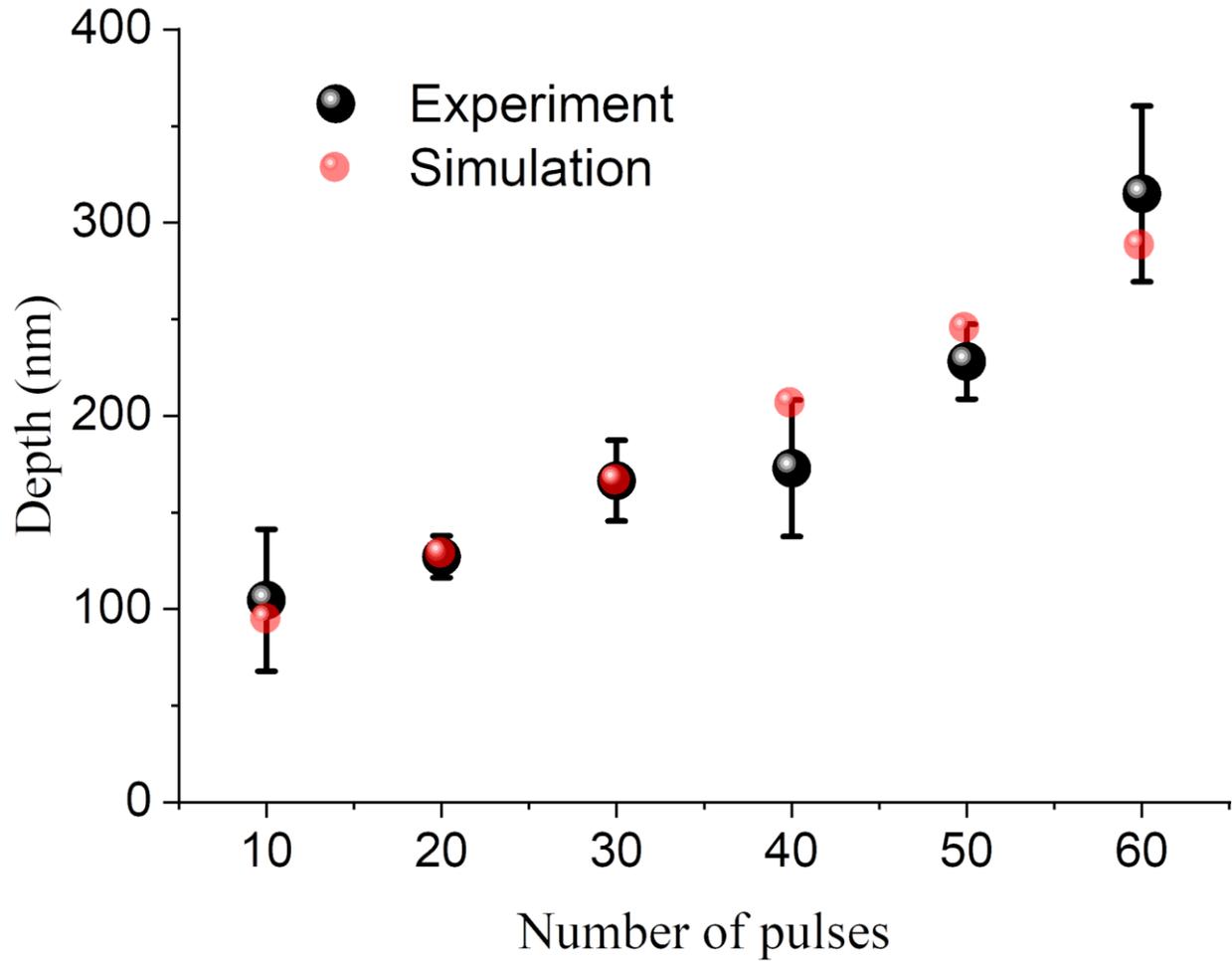

**Figure 7.** LIPSS trench maximum depths measured by AFM (black dots) and computed by full-wave multiphysics simulations (red dots). The standard deviation in the measured AFM images is calculated over at least 5 measurements per pulse count. Excellent agreement is obtained between theoretical and experimental results. The laser fluence value is always fixed to $0.23\ J/cm^2$.